\documentclass[letterpaper, 10 pt, conference]{ieeeconf}

\usepackage{hyperref}
\usepackage{multirow}
\usepackage{amsfonts}
\usepackage{epsfig}
\usepackage{amsmath}
\usepackage{amssymb}
\usepackage[nolist]{acronym}
\usepackage[english]{babel}
\usepackage{cite}
\usepackage{color}
\usepackage{stfloats}
\usepackage{psfrag}
\usepackage{algorithm}
\usepackage[export]{adjustbox}[2011/08/13]
\usepackage{algorithmic}
\usepackage{graphicx}
\usepackage{bbm}	
\usepackage{cleveref}
\usepackage{xcolor}
\usepackage{balance}
\usepackage{authblk}

\usepackage{pgfplots}
\pgfplotsset{compat=newest}
\usetikzlibrary{plotmarks}
\usetikzlibrary{arrows.meta}
\usepgfplotslibrary{patchplots}
\usepackage{grffile}
\newlength\fwidth

\usepackage{amsthm}

\let\oldref\ref
\renewcommand{\ref}[1]{(\oldref{#1})}

\IEEEoverridecommandlockouts 

\begin{document}

\begin{acronym}
\acro{ADR}{adaptive data rate}
\acro{CSS}{chirp spread spectrum}
\acro{ESP}{estimated signal power}
\acro{FSM}{finite state machine}
\acro{IoT}{internet of things}
\acro{LoRa}{long-range communication protocol}
\acro{LoRaWAN}{long-range wide-area network}
\acro{LPN}{low-power network}
\acro{LPWAN}{low-power wide-area network}
\acro{LSE}{least-square error}
\acro{MAC}{media access control}
\acro{MAVLink}{micro air vehicle link}
\acro{MAVROS}{interface between MAVLINK and ROS}
\acro{NB-IoT}{narrow-band IoT 5G standard}
\acro{PHY}{physical layer}
\acro{QGC}{QGroundControl software}
\acro{RC}{radio-controller}
\acro{ROS}{robot operating system}
\acro{RPi}{Raspberry Pi}
\acro{RSSI}{received signal strength indicator}
\acro{R-SSH}{reverse SSH}
\acro{SITL}{software in the loop}
\acro{SF}{spreading factor}
\acro{SNR}{signal-to-noise ratio}
\acro{SSH}{secure shell}
\acro{TTN}{The Things Network}
\acro{WSN}{wireless sensor network}
\end{acronym}

\title{
\LARGE \bf Drone-aided Localization in LoRa IoT Networks}

\author{Victor Delafontaine$^*$, Fabrizio Schiano$^*$, \textit{Member}, \textit{IEEE}, Giuseppe Cocco$^*$$^\dagger$, \textit{Member}, \textit{IEEE}, Alexandru Rusu$^{**}$, \\and Dario Floreano$^*$, \textit{Senior Member}, \textit{IEEE}
\thanks{$^*$Laboratory of Intelligent Systems, Ecole Polytechnique Federale de Lausanne, CH-1015 Lausanne, Switzerland}
\thanks{$^{**}$Swisscom IoT/5G Innovation Lab, CH-1015 Lausanne, Switzerland}
\thanks{$\dagger$ Information Theory and Coding Group, DTIC Department, Pompeu Fabra University (UPF), Barcelona, Spain}
\thanks{Giuseppe Cocco has received funding by the postdoctoral fellowship programme Beatriu de Pin\'{o}s, funded by the Secretary of Universities and Research (Government of Catalonia), and by the EU-H2020 RIA programme under the MSC grant agreements No. 751062 (IF) and No. 801370.}
}

\maketitle
\IEEEpeerreviewmaketitle


\begin{abstract}
Besides being part of the Internet of Things (IoT), drones can play a relevant role in it as enablers. The 3D mobility of UAVs can be exploited to improve node localization in IoT networks for, e.g., search and rescue or goods localization and tracking. One of the widespread IoT communication technologies is Long Range Wide Area Network (LoRaWAN), which allows achieving long communication distances with low power. In this work, we present a drone-aided localization system for LoRa networks in which a UAV is used to improve the estimation of a node's location initially provided by the network. We characterize the relevant parameters of the communication system and use them to develop and test a search algorithm in a realistic simulated scenario. We then move to the full implementation of a real system in which a drone is seamlessly integrated into Swisscom's LoRa network. The drone coordinates with the network with a two-way exchange of information which results in an accurate and fully autonomous localization system. The results obtained in our field tests show a ten-fold improvement in localization precision with respect to the estimation provided by the fixed network. Up to our knowledge, this is the first time a UAV is successfully integrated in a LoRa network to improve its localization accuracy.
\end{abstract}
\section*{SUPPLEMENTARY MATERIAL}
Supplementary video: 
\href{https://youtu.be/7HQ6MxwWS4w}{https://youtu.be/7HQ6MxwWS4w}

\section{Introduction} \label{sec:intro}
The advent of the \ac{IoT} represents  a major paradigm shift in communications, allowing to interconnect billions of devices, from home appliances to robotic systems \cite{mozaffari_UAV_IoT_TWC2017}\cite{iot_survey_CST_2015}. Several communication standards and protocols have been proposed in the IoT context over the years, some targeting short-range wireless communication, such as Zigbee, near field communication (NFC) or Bluetooth \cite{gubbi2013internet}, and others focusing on long-range communication. Among these, a promising technology is the \ac{LoRa} \cite{lora}. \ac{LoRa} offers both long-range communication and high energy-efficiency \cite{centenaro}. Such characteristics make it an appealing solution for battery-constrained \ac{IoT} devices intended for outdoor use and deployed in remote areas. While \ac{LoRa} technology was originally developed for communication purposes, it has been recently proven to be a valid energy-efficient alternative to other localization technologies. The localization of \ac{IoT} devices is of paramount importance for applications such as the tracking of goods or low-cost and energy-efficient emergency devices. Localization of \ac{IoT} nodes is usually tackled by including a Global Navigation Satellite System (GNSS) receiver and a wireless transmitter in the node to be localized. Despite their relatively low cost, including a GNSS receiver may be unacceptable for low-end-low-cost devices. Moreover, as mentioned in \cite{fargas2017gps}, the relatively high power consumption of the GNSS receiver greatly limits the battery lifetime.
An alternative to this is to localize the node based on the signal strength received at some reference receiving stations, that are usually assumed to be fixed  \cite{alippi, wang}. Using LoRa for this application could lead to a solution suited for low-power devices over long-range. 
\begin{figure}[t]
    \centering
    \includegraphics[width=0.9\linewidth]{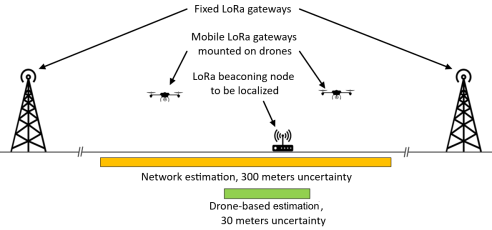}
    \caption{System setup.}
    \label{fig:systemSetup}
\end{figure}
Low-power networks (LPN) based on LoRa are already deployed in countries such as Switzerland and the Netherlands. However, the localization accuracy achieved by fixed LoRa networks is not acceptable for specific applications (see~\Cref{fig:systemSetup})\footnote{As an example, the accuracy of the localization given by Swisscom's LPN network in Switzerland is of approximately $300m$, which may not be acceptable for, e.g., localizing a person trapped under an avalanche.}. 

In the last years,  the use of UAVs in communication networks has focused mostly on performance improvement in terms of coverage or to provide additional network capacity  \cite{mozaffari_UAV_IoT_TWC2017,lora_uav_coverage,alzenad_uav_max_coverage_WCL2017}. Apart from this, UAVs can be used in the context of localization in \ac{IoT} networks, acting as 3D-mobile gateway stations, with possible applications in search-and-rescue (e.g.,  avalanches or earthquakes scenarios) or goods tracking \cite{pack_uav_for_localiz_TSMC2009}. However, so far little research has been done on the deployment of drones as mobile gateways in LoRa networks for localization applications\cite{sharma2018lorawan} and current research focuses on theoretical aspects or simulated performance, with partial or no real-world implementations.

In this work, we study the use of UAVs as mobile gateways to improve localization in \ac{LoRa} IoT networks. We start by experimentally deriving a statistical characterization of the \ac{LoRa} signal. The statistical model is then used to develop a search algorithm with one drone using realistic propagation channel and UAV dynamics. The algorithm is then extended to three drones in order to decrease the mission duration while achieving the same precision. The simulations are first implemented in MATLAB and then through Software In The Loop (SITL) simulations to test the feasibility of our approach. The proposed system has then been implemented using a real drone as mobile gateway and integrating it within the existing Swisscom \ac{LoRa} network. Our tests show that a dramatic improvement of more than one order of magnitude in localization precision (from 300m to 12m) can be obtained already with a single drone. Although using drones for target localization has already been proposed in literature \cite{drone_assist_loc_TVT}\cite{uav_aided_ICC}\cite{globecom_reference}, little work has been done to demonstrate its practical feasibility, including the interface with an existing communication network. Up to our knowledge, this is the first time that the feasibility of a drone-aided localization system based on LoRa technology has been proven.

The rest of the paper is structured as follows. In \Cref{sec:state} the state of the art is revised. In \Cref{sec:system} we introduce the system model. \Cref{sec:carac} contains the statistical characterization of the signal between the beaconing node and the mobile gateways. The search algorithm for one and three UAVs based on the statistical model  is presented in \Cref{sec:matlab}. \Cref{sec:matlab2} contains the simulation results while the field test results are reported in \Cref{sec:exp_results}. Finally, \Cref{sec:conc} describes the conclusions and future works.

\section{Related Work} \label{sec:state}
LoRa is a proprietary data communication technology developed for IoT, which defines the so called \emph{LoRaPHY} physical layer. LoRaPHY is proprietary and no official documentation is available, although some of the key parameters can be derived empirically \cite{aloys}\footnote{\href{https://revspace.nl/DecodingLora}{https://revspace.nl/DecodingLora}}. Unlike the physical layer, the medium access control (MAC) layer usually used with LoRa, called LoRa Wide-Area Network (LoRaWAN), is an open standard and has been developed by the LoRa Alliance \cite{lora_alliance}.
One of the reasons for LoRa popularity and widespread utilization is the fact that it naturally fits key IoT applications characterized by energy-constrained nodes scattered over a wide geographical area. For instance, a LoRaWAN network has a tested range of up to two kilometers in urban areas, capable of servicing 200k nodes in 100 square kilometers using 30 gateways \cite{centenaro}. Two AA batteries can give a device a battery life of two years in the case of small periodic messages \cite{bor2016lora, adelantado}.

Among the methods for node localization based on communication networks, those most related to the present work are multilateration and fingerprinting \cite{yang,wang}\footnote{Although a time-based approach as in GPS or Galileo systems can in principle work, this requires very high clock accuracy, which is typically not available in IoT networks or requires costly network upgrades \cite{csem_time_based}.}. Fingerprinting compares an existing map of signal data obtained experimentally and the received data. It has a generally good accuracy, but is time-consuming, labor-intensive and vulnerable to environmental dynamics, which makes it unsuited for our scope \cite{yang}.
In multilateration methods a curve linking signal strength and distance is obtained experimentally \cite{alippi}. Measurements of the signal strength at different locations are then used to  obtain an estimation of the position. This method can be applied with different communication technologies \cite{alippi} such as Bluetooth \cite{wang} or WiFi.

The agility, speed and manoeuvrability of UAVs make them suited for many IoT applications \cite{fotouhi}. Their main limitation is the battery lifetime that restricts the applications to relatively short time periods. 
Recently, the deployment of a local WiFi network using fixed-wing drones has been proposed~\cite{rosati2013testbed}.
A combined application of LoRa and UAVs was proposed for surveillance application in \cite{sharma2018lorawan}. In the paper the authors study the possibility of deploying on-demand nodes embedded on a UAV for improving security in intelligent transportation systems (ITS). Unlike the present paper, in \cite{sharma2018lorawan} LoRa is used with a communication purpose rather than a localization one.
One of the main advantages in using drones as mobile base stations/gateways is that they can be deployed on-demand and increasing their number can increase the efficiency of the system \cite{hayes2001swarm, madhavan2002distributed}. In \cite{hayes2001swarm} it was shown that a swarm deployed over a large area can localize a source faster than a single robot. Applying an extended Kalman filter in a multiple robots setting can also allow to determine their relative positions \cite{madhavan2002distributed,schiano2018dynamic}.

\section{System Architecture} \label{sec:system}
\subsection{Baseline System}
Let us consider a LoRaWAN network composed of a LoRa IoT user node to be localized, a set of network gateways and a LoRaWAN server. LoRaWAN was originally conceived as a communication network meant for user nodes to send data messages to the gateways. Here, we use LoRaWAN to localize the user node by measuring the power received at the gateway stations, as detailed in the following. Upon receiving a message from the node, the gateway adds a layer of metadata before forwarding it to the network server. Such metadata contains, among other information, an indicator of the received signal strength, a timestamp indicating the reception time and the gateway ID. The message with the metadata is passed on to the LoRaWAN server which then routes it to a backhaul server that stores the messages in a database. 
Note that, due to the broadcast nature of the wireless medium and the typical LoRa gateways locations, the same message is usually received by more than one gateway. In such case the gateways (the location of which is precisely known) act as sensors measuring the received signal power level. At the backhaul server, the gateway ID and the received power level indicator are extracted from each of the gateways' messages and merged using the known location of the gateways. In this way an estimation of the node's location can be obtained.

\subsection{Considered Setup}
The accuracy of the network estimation is typically in the order of $300$m, which might not be sufficient for time-critical applications. To improve the estimation, we propose to use a UAV as a mobile gateway. The UAV carries a LoRa gateway, which receives the messages transmitted by the node, a 4G communication module, that relays the message with metadata to the LoRa server and receives commands from the network, and a GPS receiver, which is used to localize the UAV and allow triangulation. Upon receiving the data from the drone, the LoRa server fuses these with the data obtained from the fixed gateways in order to improve the initial network estimation. The system has been designed so that the UAV is seamlessly integrated in the LoRa network, i.e., from the perspective of the LoRa server it is seen as an additional gateway, which allows for smooth internetworking and data processing, while retaining all benefits of a mobile gateway which can be deployed on-demand where and when the need arises. Specifically, starting from the original network estimation, the locations where the drone will perform the measurements are calculated by the backhaul server and transmitted to the drone as waypoints. Thanks to the fact that the UAV can get closer to the beaconing node than the fixed LoRa gateways and to the spatial diversity achieved through mobility, the measurement precision increases and, consequently, the localization accuracy improves.

The specific LoRa modules we used in the implementation part of the present work are a Gimasi Tuino One with a LoRa module for the node, and the Multitech MultiConnect Conduit, equipped with a LoRa module as well as built-in mobile internet connectivity, for the on-board gateway.

\section{Signal Characterization and search algorithms design} \label{sec:carac}
In order to derive the distance between the transmitting node and the receiving gateway from the received signal power, the following expression (in logarithmic scale) for the received power can be used \cite{maral_satellite}:
\begin{equation}
P_{\text{RX}} = P_{\text{TX}} + G_{\text{Tx}} + G_{\text{Rx}} - L_{\text{FS}} - L_{\text{O}} + N \text{ [dBW]}
\label{eq:link_bud_noise}
\end{equation}
where $L_{\text{FS}} \triangleq \alpha\log_{10}\left({4\pi R}/{\lambda}\right)$,
$R$ being the transmitter-receiver distance and $\lambda$ the communication wavelength, $P_{\text{TX}}$ is the transmitter's power while $L_{\text{O}}$ takes into account other losses such as those between the amplifier and the antenna in both terminals. The terms $G_{\text{Tx}}$ and $G_{\text{Rx}}$ in \Cref{eq:link_bud_noise} represent the transmitter's and receiver's antenna gains, respectively. Such values depend on the specific antenna considered and are, in general, a function of the direction from which each terminal is seen from the other.
The coefficient $\alpha$ in the definition of $L_{\text{FS}}$ takes into account the propagation environment and, in case of a free-space propagation, is equal to $2$. If all parameters are known and assuming the received power is measured at the gateway, an estimate of the distance $R$ can be obtained inverting \Cref{eq:link_bud_noise}. However, the presence of the multi-path propagation noise
 $N$, makes the estimation of the distance more challenging. Furthermore, unlike the free-space propagation case, the parameter $\alpha$ in the definition of $L_{\text{FS}}$ assumes a value which can be different (usually higher) than $2$.
Since not all of the parameters are available from hardware data-sheets or from LoRa documentation (which is not fully public), we carried out several experiments to estimate them in a controlled environment in which both the distance and the relative antenna orientation between the node and the gateway are known. In order to have a clear  view of the parameters to be estimated, we rewrite \Cref{eq:link_bud_noise} as:
\begin{equation}\label{eq:link_bud_ab_noise}
P_{\text{RX}} = \log_{10}\left(R^{1/b}\right) - \log_{10}(a) + G_{\text{Tx}} + G_R  \text{ [dBW]} + N,
\end{equation}
where $a$ and $b$ are the unknown parameters that account for transmit power, hardware losses and propagation losses. 

\subsection{Received Power}
The received power is measured at the gateway in the form of  received signal strength indicator (RSSI). From this and from the signal to noise ratio (SNR), also available at the receiver, the Estimated Signal Power (ESP) as\footnote{Typical ranges for the metadata that we encountered are $[-100,-30]$ dBm for the RSSI and ESP and $[5,10]$ for the SNR.} can be computed as:
\begin{equation}
    \label{eq:rssi_esp}
    ESP = RSSI - 10 \log \left( 1 + 10^{-(SNR/10)}\right)
\end{equation}
Although RSSI and ESP have different definitions (RSSI also includes the thermal noise at the receiver, ESP does not), either of them can be used in \Cref{eq:link_bud_ab_noise} to estimate the unknown parameters.
In the following sections we characterize the unknown parameters in \Cref{eq:link_bud_ab_noise}, namely  the antenna gain, the noise statistics and the $a$ and $b$ coefficients.
\subsection{Antenna Gain Characterization} \label{sec:antenna}
The antenna radiation pattern has been experimentally characterized.
Measurements were made using different relative positions of the receiver with respect to the transmitter. During the measurements, the distance between the two devices was kept constant to isolate the impact of the antenna radiation pattern. 
The resulting radiation pattern is omnidirectional in the horizontal plane while is directive in the vertical plane, with blind spots at the poles.
A linear fit for the angles under $60$ degrees resulted in the overall gain due to both antennas, given in \Cref{eq:att}, where the angle $\theta$ is in degrees. The coefficients' values are  $a_{ang}=0.5667$ and $b_{ang}=1.38$. These values are obtained using the ESP, and the corresponding values for the RSSI are close enough to be considered as equal.
\begin{align}
    \label{eq:att}
    G_{\text{tot}} = G_{Tx} + G_{Rx} = -( a_{ang} \cdot \theta + b_{ang})
\end{align}

\subsection{Noise characterization} \label{sec:noise_carac}
Measurements were made with the transmitter (LoRa node) located at the height of $10$ meters from the ground, while the receiver (LoRa gateway) was placed on the ground at distances in the range between $10$ and $150$ meters. The measurements were made in a location characterized by tall buildings (around $20$ meters) with a clear line-of-sight between the devices.  Several measures were taken for each distance to average out the effect of noise. {Note that the propagation environment of the tests has in general an impact on the model and thus on the precision of the distance estimation during the operations. Since the final tests with a real drone and LoRa node to be located were performed in a different propagation environment (open field), the measures suffer from a model mismatch. However, as described in \Cref{sec:exp_results}, our approach is robust under such mismatch, allowing us to achieve a ten-fold improvement with respect to the fixed LoRa network.} This is of great practical relevance since the operating environment could often not be known in advance and can differ significantly from the one where the propagation model was derived. We repeated the characterization by using data collected on an open field, which further improved the results. A full characterization of all main propagation scenarios (e.g., urban, suburban, hilly terrain) is out of the scope of this work (see \cite{lora_uav_coverage} for further details).
No significant correlation was observed between distance and noise's standard deviation, which was around 2 dBm. A normal fit on the signal expressed in dB has been used, which is in line with the widely adopted lognormal fading model. The standard deviation used in the simulation is slightly larger than the one recorded (2.5 instead of 2) to account for the vibrations and movements of the drone, which depend on the operational conditions (e.g., winds) that cannot be known in advance.
\subsection{Path loss characterization} \label{sec:exponential}
\begin{figure}[H]
    \centering
    \includegraphics[width=0.9\linewidth, center]{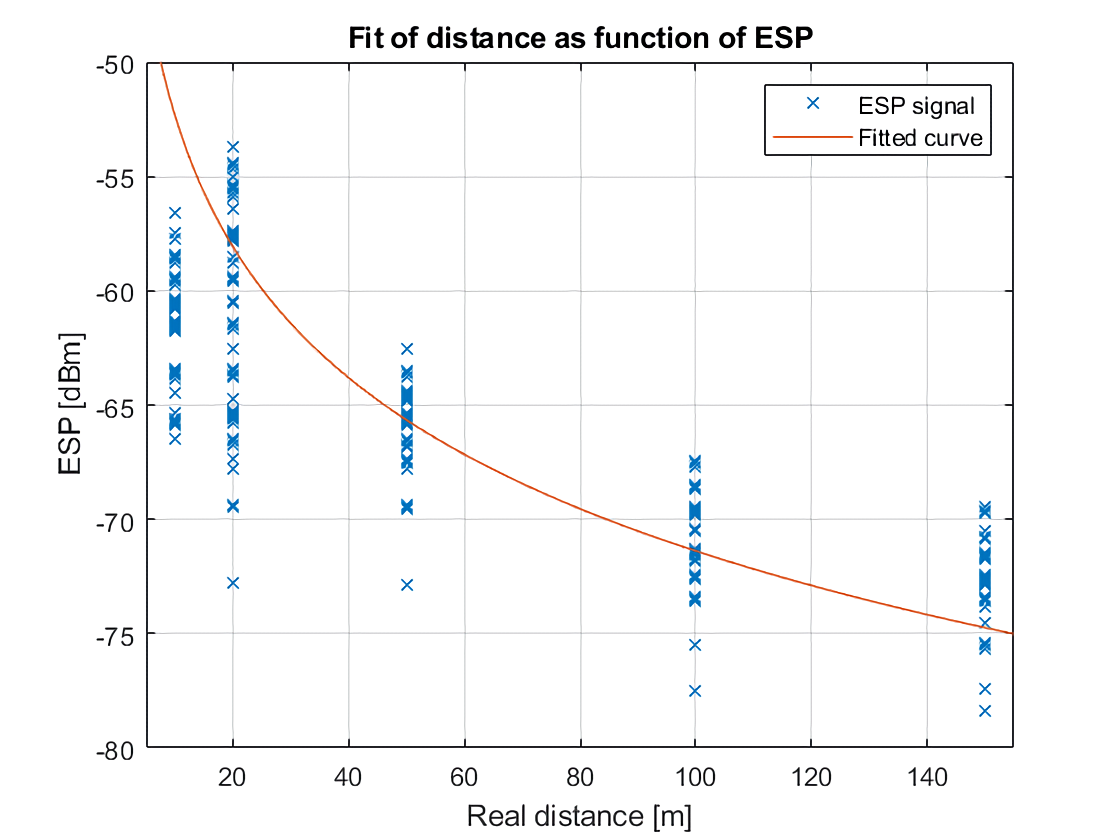}
    \caption{Exponential curve-fitting of ESP versus distance.}
   \label{fig:fit_dESP}
\end{figure}

Let us now consider the $a$ and $b$ parameters in \Cref{eq:link_bud_ab_noise}. The measured data were fitted according to the exponential curve shown in \Cref{fig:fit_dESP}. 
Since, as can be seen from \Cref{eq:link_bud_ab_noise}, the received power also depends on the antenna gain, we compensated for it using the model derived in subsection \ref{sec:antenna}.
The estimated $a$ and $b$ coefficients using ESP and RSSI of \Cref{eq:link_bud_ab_noise} are shown in \Cref{tab:results_ab}.

\begin{table}[H]
    \centering
    \caption{Free-space path loss exponential coefficients}
    \begin{tabular}{c|c|c|}
        \cline{2-3}
        & {$a$} & {$b$} \\ \hline
        \multicolumn{1}{|l|}{ESP}  & 0.1973 & -0.0902 \\ \hline
        \multicolumn{1}{|l|}{RSSI} & 0.2189 & -0.0894 \\ \hline
    \end{tabular}
    \label{tab:results_ab}
\end{table}

\subsection{Search Algorithm} \label{sec:matlab}
The propagation model presented previously in this section has been used to develop different search algorithms and compare them to find a suitable one to be implemented on the UAV.
 Notice that, since the focus of this work is not in the algorithm itself, we focused on a low complexity solution which is also robust against signal noise. Potential improvements are discussed in~\Cref{sec:conc}.
 
We considered two setups: one with one drone and one with three drones. In each of them we compared two different approaches: in one the drones perform continuous measurements (later on transmitted to the server for multilateration) as they move along a predetermined trajectory while in the second approach they make several measurements while hovering in the same location before moving to the next one, in order to average out the noise. 
The first approach removes the need to hover for extended periods to collect data, at the cost of noisier measurements. \footnote{Another advantage of making continuous measurements is that it could also be used for drones without hovering capabilities such as fixed-wing drones}. A greedy algorithm has been considered in both cases. At each iteration, measurements are made on a circle centered around the last estimated node location. The circle radius is decreased at each iteration. The circle radius is determined numerically, as explained in next section. The choice of a circular sampling curve was motivated by the circular symmetry of the problem which, in turn, comes from the symmetric antenna radiation pattern.

\section{Simulation Results} \label{sec:matlab2}
In order to  get an insight on the required flight time and estimation accuracy, the search algorithms were tested first through MATLAB simulations and then in SITL simulations using the Gazebo simulator, ROS, and the PX4 firmware. This allowed us to test our solutions at increasingly realistic steps before moving to a real-world implementation (see~\Cref{sec:exp_results}). {Thanks to these simulations we were able to tune different parameters, such as the measurement circle radius, that are discussed below.}
In the following subsections, four different simulation setups are considered. 
\subsection{One UAV, discrete measurement method} \label{sec:1drone_tri}
\begin{figure}[t]
    \centering
    \includegraphics[width=0.87\linewidth]{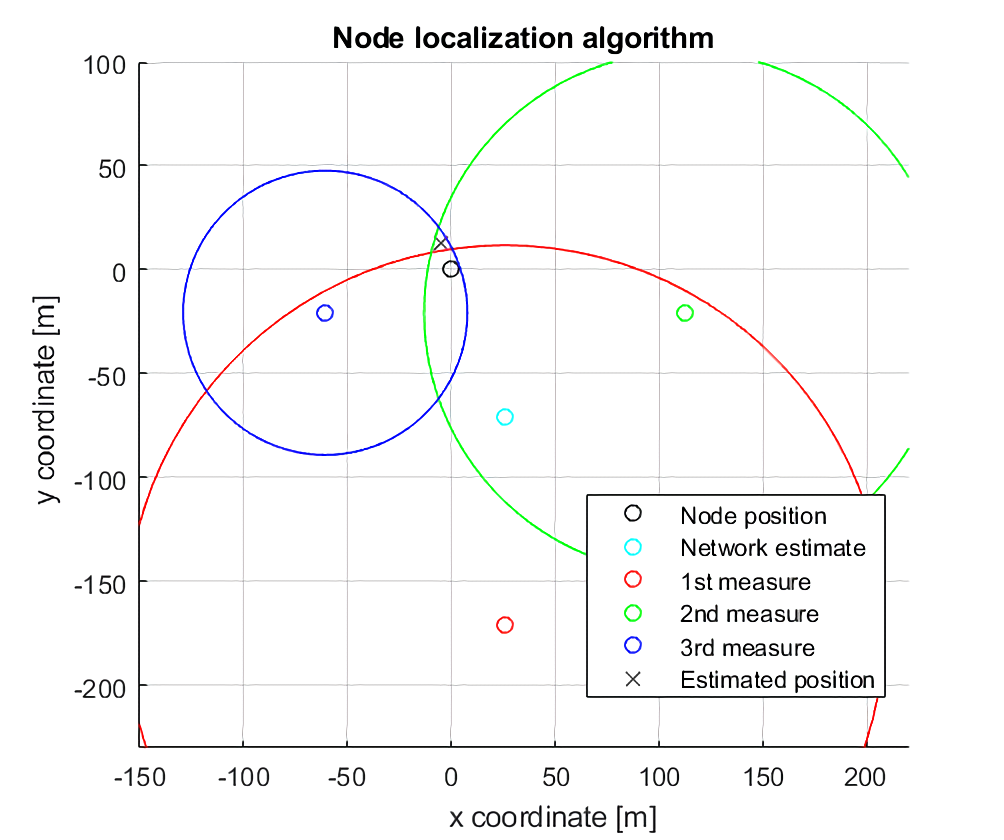}
    \caption{Overview  of the discrete measurement method. At each measurement iteration, the radius of the circle where the measurement points are located decreases.}
    \label{fig:1drone_tri}
\end{figure}

{\Cref{fig:1drone_tri} shows an example of the \emph{discrete method}}. 
The network estimation of the node position is the starting point of our simulations. Our algorithm starts by setting three locations on a circle centered in such estimation. The UAV will then visit these three positions and perform LoRa signal measurements. The radius of the circles (red, blue, green) pictured in \Cref{fig:1drone_tri} correspond to the distance estimated using the power-distance model in \Cref{sec:carac}. The radius of the circle on which the three waypoints are defined was set considering the initial network accuracy and optimized numerically tacking the signal model into account. After this first iteration, a new estimation of the node location is obtained. The UAV starts a new measurement iteration, in which the initial network estimate is replaced with the new one. The radius of the circle where the three measurements are made is determined numerically taking the noise variance into account, decreasing at each iteration {to improve accuracy}.
 With the continuous method an average precision of $5.4$m was achieved in a time between $4$ and $6$ minutes. 
\subsection{One UAV, continuous measurement method} \label{sec:1drone_cont}
\begin{figure}[t]
    \centering
    \includegraphics[width=0.9\linewidth]{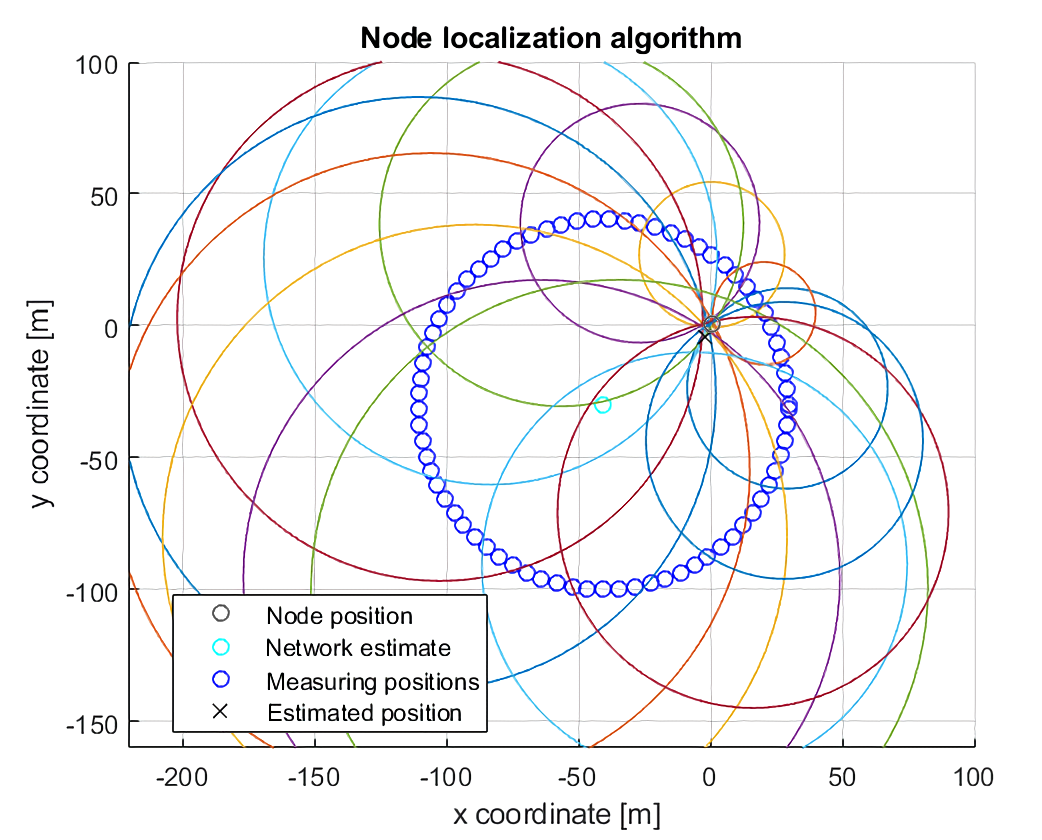}
    \caption{Overview of  the continuous measurement method.}
    \label{fig:1drone_continuous}
\end{figure}

\Cref{fig:1drone_continuous} describes the \emph{continuous} measurement method with one UAV. Similarly to the discrete version, the radius of the circle on which the measurements are done is determined numerically and decreases at each measurement round. 
The precision reached was $8.2$ m in a time between $7$ and $8$ minutes. Here, unlike the discrete method, measurements are made while the drone is moving rather than hovering in a fixed position. The UAV was set to fly at lower speed than in the discrete method (hence the longer time) to reduce the combined effect of the on-board antenna tilt due to drone's pitch and the antenna's directivity\footnote{An increase in the drones' speed would reduce the acquisition time, but also make either the measurements more noisy or increase the complexity of the triangulation algorithm to take, e.g., drone's pitch into account.}. An advantage of this method over the discrete one is that the server can do the multilateration before the UAV reaches its final measurement location. This could be of vital importance in search-and-rescue scenarios where a first rough position estimation must be provided as fast as possible.
\subsection{Three UAVs, discrete measurement method} \label{sec:3drone_tri}
The same method described in~\Cref{sec:1drone_tri} has been tested using three drones. Only one of the three measurement positions is assigned to each drone. Since each UAV needs to visit only one point, the time to accomplish the mission is smaller while  the same accuracy is achieved. We performed simulations with the same parameters of~\Cref{sec:1drone_tri} and we had the same final precision, as expected, but in half the time\footnote{Note that the time is not one third of the single-drone case due to the initial time needed to get to the initial measurement position.} (between $2$ and $3$ minutes).
Since this method requires a shorter flight time, this can be traded with increased accuracy in two ways.
The first one is to make more measurements at each measuring position. This allows to average out the signal noise and lead to better distance estimates. The second option is to add more trilateration iterations centered on the best estimate provided at the end of each iteration. A comparison of the different methods is shown in \Cref{tab:3drone_tri}. As expected, the larger the number of iterations or measurements, the better the result. 
\begin{table}[H]
    \centering
    \caption{Change in precision and flight time when increasing the number of iterations (or number of measurements) at each measuring position.}
    \begin{tabular}{|c|c|c|c|}
        \hline
        \textbf{Iterations} & \textbf{Measurements} & \textbf{Precision {[}m{]}} & \textbf{Flight Time {[}s{]}} \\ \hline
        2 & 2 & 4.1 (ref) & 175 (ref) \\ \hline
        2 & 4 & 2.6 (-36\%) & 268 (+53\%) \\ \hline
        3 & 2 & 3.2 (-22\%) & 225 (+28\%) \\ \hline
        3 & 4 & 2.3 (-44\%) & 360 (+105\%) \\ \hline
    \end{tabular}
    \label{tab:3drone_tri}
\end{table}
\subsection{Three UAVs, continuous measurement method} \label{sec:3drone_cont}
Similarly to what described in~\Cref{sec:3drone_tri}, we can use the method described in~\Cref{sec:1drone_cont} with three UAVs. 
In this case each UAV follows a circular trajectory which is centred around its measuring positions. The three measuring positions are located on a circle centered on the network estimation.
The algorithm performs two main iterations. In the first one the radius of the circle centered in the network estimate was set to  $80$m while the orbiting radius of each UAV was set at $70$m. In the second iteration the radius of the circle centered on the estimate of the first iteration is set to $50$m while the orbiting radius of each UAV was set to $30$m.
With these parameters we reached an average precision around $8.5$ m in $5$ minutes and $20$ seconds of flight (the average is taken across simulations).

\section{Experimental Results} \label{sec:exp_results}
To prove the feasibility and the validity of our approach we implemented the system described so far using one UAV, shown in the attached video, and the algorithms presented in \Cref{sec:1drone_tri,sec:1drone_cont}. 
The drone was equipped with a PixHawk flight controller (running the PX4 autopilot), a GPS module, a Raspberry Pi 3B+ onboard computer (running ROS and MAVROS), and a MultiTech MultiConnect Conduit LoRa gateway. The connection between the RPi and the Swisscom server was established through a Huawei E3372 USB modem also mounted onboard. 
In the experiments the server acted as coordinator, receiving the LoRa messages transmitted from the emitting node and forwarded (with the additional metadata) by the different gateways, and storing them in a database. In addition to the fixed gateways, the UAV also sends information on its coordinates and ancillary parameters. We differentiated the two types of messages based on unique identifiers for the UAV and for the LoRa fixed gateways. The result is a dataset, containing both position of each receiver and measured signal strength information of the LoRa node's signal, on which multilateration can be done. Then, the server computes the waypoints based on the latest position estimation according to the algorithm prsented in previous section.
A GUI was  created using the Google Maps API to have a real-time feedback and control on the UAV flight as well as the position and waypoints computed by the server  (see attached video).

We performed different flight tests, obtaining a precision between $16$m and $48$m with only one measurement iteration and one UAV and using the parameters presented in \Cref{sec:1drone_tri,sec:1drone_cont}. The continuous method returned better results (precision average of $27$m) than the discrete one (precision average of $40$m). The reason for this is that the LoRa transmission has a time period of $4$s. During a continuous flight of around two minutes, a total of $30$ LoRa messages are emitted by the beacon, but the average number of multilateration datapoints reeceived was around $20$, i.e.,  about one third of the total datapoints were lost due to synchronization issues. 
In the discrete case the UAV should hover for around $40$s in each measurement location to avoid such issues, which would double the flight time for each iteration.
While the obtained precision is already a ten-fold improvement over the one obtained from the LPN network, the accuracy can be further enhanced. In the first place, in the experiments we used the exponential fit described in~\Cref{sec:carac}, which was obtained from data collected in a urban scenario  while the  experiments were done in an open field. The statistical mismatch is responsible for part of the residual localization error. To confirm this,  we improved our fit by doing new measurements directly from the drone and in an open field (in a different location from the one used in the tests). In addition, the antenna gain described in~\Cref{sec:antenna} has also been taken into account. Furthermore, a linear fit rather than an exponential fit in the distance-strength curves has been performed, leading to a lower error. 
A new flight test was done using the computation of the precision based on the new propagation model. The localization accuracy  using the discrete method had an improvement of $83\%$ (from $48.4$ to $8$ meters) and $70\%$ for the continuous one (from $28.7$ to $8.6$ meters). This shows that having a signal model adapted to the flight environment can further improve the localization accuracy of the proposed system.

\section{Conclusions and future work} \label{sec:conc}
We designed, implemented and tested a drone-aided system to improve localization in LoRa IoT networks. We performed a measurement campaign to statistically characterize the dependency of the distance on the received power. The model was then used to design a search algorithm involving one and three drones, later on tested through MATLAB and SITL simulations. We then moved to a full implementation of the system. This involved the realization of a quadrotor capable of carrying the required hardware, the implementation of a software interface to allow the drone to interact with the Swisscom LoRa network located in Switzerland, the extraction of metadata at the LoRa server and the closed-loop communication between the central server, where the designed algorithm was running, and the drone. We performed a flight test campaigns in which a LoRa node was localized using the developed UAV and the Swisscom LoRaWAN network. Our test results showed a ten-fold improvement in the localization precision with respect to the LoRaWAN-only system from $300$ m to around $30$ m in the worst case scenario, i.e., using a propagation model fitted with data from a different propagation environment than the test one. In the best case scenario in which a similar environment was used for the model fitting and the test, a precision of around $12$m was achieved. This encouraging result shows the great potential of UAVs as mobile gateways to significantly improve the localization accuracy of existing fixed networks. Potential applications of the proposed scheme range from search and rescue missions to on-demand goods tracking and localization. Up to our knowledge, this is the first time a UAV has been successfully tested as a mobile gateway to improve localization in LoRA networks.

As future research directions, an interesting possibility is to use an extended Kalman filter to further improve the localization accuracy \cite{madhavan2002distributed, giannitrapani2011comparison}. This could allow to include the additional information from the antenna radiation diagram (e.g., the presence of a blind spot) in the search algorithm. From the perspective of UAV path and waypoints calculation, this could be implemented using real-time trajectory optimization for localization as in \cite{ponda2009trajectory}. 
Also, to make the localization more robust it would be interesting to make the localization algorithm dependent on the environment in which the system is deployed.
Finally, in the multi-drone scenario, machine-to-machine (M2M) communication between the UAVs (possibly using the same LoRa module to reduce on-board weight) could help in case communication with the server is lost.

\balance
\bibliographystyle{IEEEtran}
\bibliography{IEEEabrv,99_biblio}

\end{document}